# Convergent Fabrication of a Nanoporous Two-Dimensional Carbon Network from an Aldol Condensation on Metal Surfaces

John Landers,[1,2] Frédéric Chérioux,[3*] Maurizio De Santis,[1,2] Nedjma Bendiab,[1,2] Simon Lamare,[3] Laurence Magaud,[1,2] Johann Coraux[1,2*]

[1]*Univ. Grenoble Alpes, Inst NEEL, F-38042 Grenoble, France*
[2]*CNRS, Inst NEEL, F-38042 Grenoble, France*
[3]*Institut FEMTO-ST, Université de Franche-Comté, CNRS, ENSMM, 32 Avenue de l'Observatoire, F-25044 Besançon, France*

Email:
johann.coraux@neel.cnrs.fr
frederic.cherioux@femto-st.fr

**Abstract**

We report a convergent surface polymerization reaction scheme on Au(111), based on a triple aldol condensation, yielding a carbon-rich, covalent nanoporous two-dimensional network. The reaction is not self-poisoning and proceeds up to a full surface coverage. The deposited precursor molecules 1,3,5-tri(4'-acetylphenyl) first form supramolecular assemblies that are converted to the porous covalent network upon heating. The formation and structure of the network and of the intermediate steps are studied with scanning tunneling microscopy, Raman spectroscopy and density functional theory.

**Article's reference: 2D Materials, 2014, 1, 034005 – DOI:10.1088/2053-1583/1/3/034005**

**Introduction**

The family of two-dimensional (2D) materials is a rapidly growing one,[1,2] which emerged with the study of graphene. Tailoring the structure of these materials allows engineering their properties. Striking examples are the change in the topology of the electronic band structure of graphene as a function of the number of layers,[3,4] the occurrence of electronic resonances in graphene islands under strong compressive strain,[5,6] and the confinement in graphene nanoislands,[7-9] and nanoribbons.[10] Hollow versions of graphene offer new degrees of freedom for structure-engineering of the properties. Examples of such hollow 2D materials are graphynes,[11] graphdiynes,[12] and graphene antidot lattices.[13] The first two consist of low-density carbon atomic lattices with sub-nanometer pores, and have been predicted to exhibit exciting electronic properties, arising from multiple and/or anisotropic Dirac cones in their electronic band structure.[14] The latter has been much discussed in the literature, as it is predicted to yield spin qubits,[13] to open sizable band-gaps in an otherwise gapless graphene,[15] and to host dispersion-less electronic bands.[16] Experimental realization of graphene antidot lattices has been mostly achieved with the help of top-down approaches relying on lithography performed on plain graphene sheets.[17,18] An alternative approach would consist of a controlled assembly of well-chosen molecular blocks. Accordingly, covalent nanoporous networks, which are actually graphene antidot lattices with ultimately thin pore walls, have been prepared by interfacial Ullmann coupling reactions at metallic surfaces, consisting of a 2D polymerization of halogenated aromatic monomers.[19] Recently, other 2D reaction schemes yielding nanoporous networks have been discovered,[20,21] which also have the potential to yield conjugated nanoporous networks with electronic properties like those of graphene antidot lattices. In addition to being an ultimately precise version of the antidot lattice, these systems are also of great interest for the prospect of nanosieving or nanosensing, enabled by functionalization of the pores.

An ongoing effort has focused on the development of surface chemistry schemes capable of delivering nanoporous 2D covalent networks. The synthesis of purely carbon networks has been demonstrated recently by a few reactions, for instance the above-mentioned Ullmann coupling,[19] homo-coupling of alkynes,[20] and diyne cyclotrimerization.[21] One promising alternative would be to employ a convergent pathway, by which several molecular monomers react together, promoted by strong energy gains, to form for instance additional aromatic rings.[22] To date such convergent methods have not yet been exploited for achieving carbon-rich networks. We report such a reaction, inspired by previous works formerly implemented in solution by one of us, to achieve fully-conjugated hyper-branched dendrimers.[23-25] The reaction proceeds by a triple aldol condensation reaction between three acetyl groups, and is demonstrated here on Au(111). Unlike in previous reports of aldol condensation for the polymerization of methyl pyruvate on Pt(111),[26-28] which only yielded disconnected agglomerates, we observe extended 2D covalent nanoporous networks. The reaction is not self-poisoning and proceeds up to a full surface coverage. The only by-product is water, which is readily desorbed from the surface under the explored conditions. Using scanning tunneling microscopy (STM), Raman spectroscopy and density functional theory (DFT) calculations, we identify intermediate states in the growth process. The network is stable up to 500°C under ultra-high vacuum (UHV) and when exposed to ambient conditions, the carbon backbone remains intact.

**Results and discussion**

We have used 1,3,5-tri(4'-acetylphenyl)benzene (see Scheme 1), referred to in the following as TriAc, as the molecular precursor for the reaction. TriAc molecules were deposited under UHV onto the herringbone-reconstructed (111) surface of Au, between room temperature and 400°C. The molecules form domains which nucleate between the ridges (partial dislocations[29]) of the herringbone reconstruction of Au(111), where adsorption is expected to be more favorable (Figure S1 in the Supporting Information).[30] The molecules form compact supramolecular networks, giving rise to spotty electron diffraction patterns (Figure S2 in the Supporting Information), and whose degree of order and symmetry vary as a function of the surface coverage. Figure 1a shows one of these supramolecular networks, obtained uniformly across the sample surface, consisting of ordered domains typically extending over 50 nm$^2$, obtained for a TriAc dose prior to the formation of the second molecular layer, and consisting of a dense arrangement of interpenetrated molecules. In this arrangement, each oxygen atom of a TriAc interacts with one hydrogen atom on the $C_2$ ortho position of the lateral benzene ring on a neighboring TriAc molecule (see zoomed-in view of Figure 1b). The STM analysis reveals a 0.2 nm distance between these H and O atoms, which is typical of a hydrogen bond between acetyl and phenyl groups.[31,32] Other types of H-bond stabilized supramolecular networks were found, some of which are seen in the phase diagram of Figure 2 and reported in Figure S3 in the Supporting Information. Noteworthy, the ridges of the Au(111) herringbone reconstruction are still observed in the presence of the supramolecular networks, for instance seen in the form of a brighter line of TriAcs in the diagonal running from bottom left to top right in Figure 1a. No substantial decrease of the surface coverage is observed upon annealing for a few minutes between room temperature and 200°C, unlike other molecules of similar size (see, e.g., Ref. 33), i.e. the bonding of the TriAcs with Au(111) is substantial. Actually, ultraviolet photoelectron spectroscopy of oxygen-containing molecules on Au(111) point to a soft chemisorption,[34] a likely situation in the present case as well. Depending on the preparation of the tip, STM reveals a non-uniform conductance of the molecules (inset of Figure 1a), pointing to variations of the electronic density of states (DOS) (Figure S3 and S4 in the Supporting Information). In the following, unless specified, all experiments consist of TriAc deposition at 200°C for a few minutes, a temperature at which highly ordered supramolecular networks still form without the TriAc molecules reacting with one another, followed by annealing at higher temperatures.

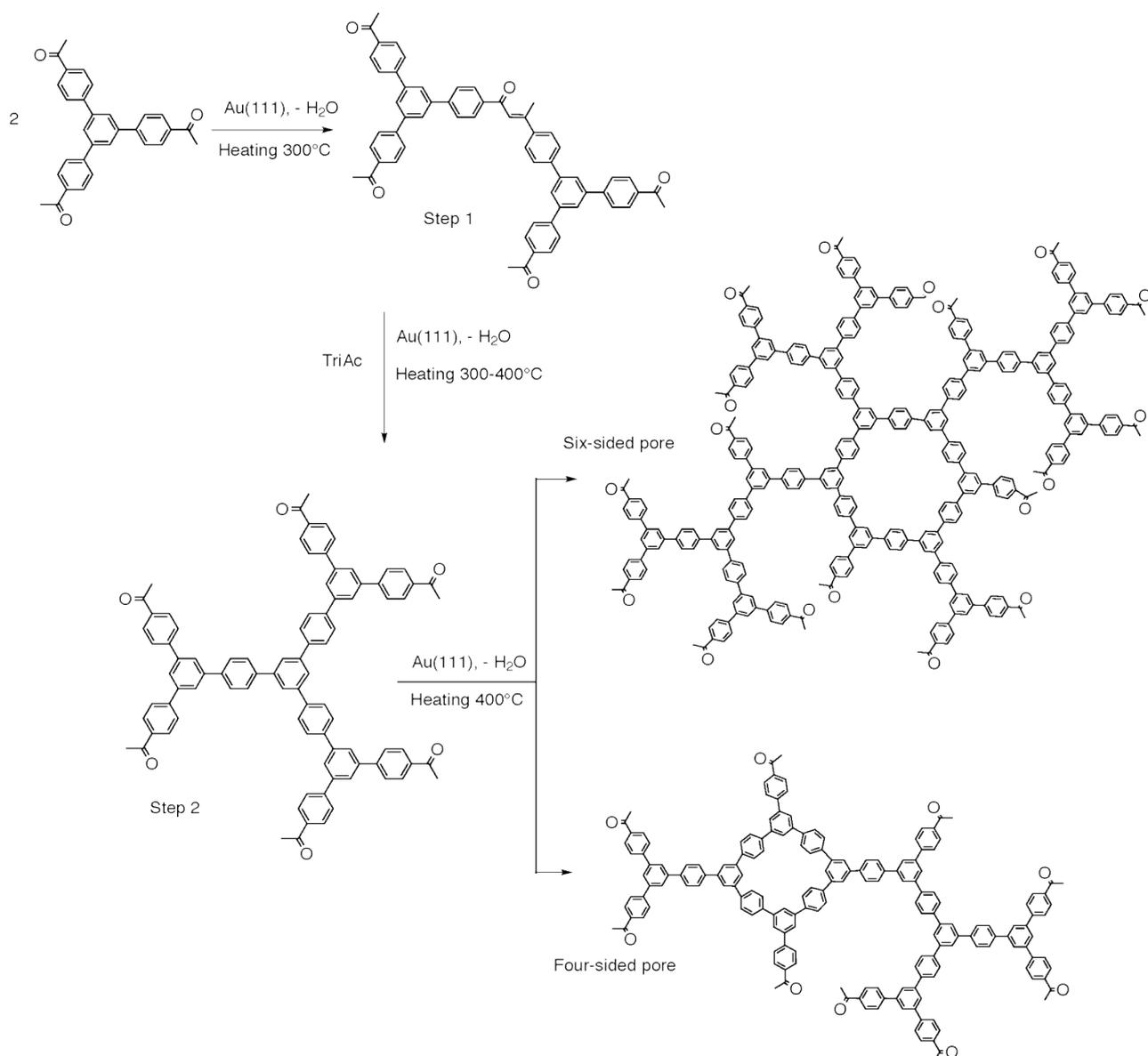

***Scheme 1:*** *Aldol condensation surface reaction. Upon moderate heating (300°C) two TriAc molecules react together on Au(111) to form an intermediate dimer while releasing a water molecule (Step 1). At higher temperature (>300°C), the convergent triple aldol condensation reaction sets in, which yields a new benzene ring (Step 2). At higher temperatures (>400°C), the polymerization reaction sets in, yielding a 2D nanoporous polymer, prominently composed of four- and six-sided pores, involving at least one new benzene ring each (here three and two extra benzene rings are formed for six- and four-sided pores).*

When longer annealing times are employed at 200°C, consisting of hours instead of a brief period of a few minutes, a progressive disorder begins to develop in the supramolecular network. This disorder is associated with the TriAc molecules forming head-to-head dimers, as observed in STM images (Figure S5 in the Supporting Information). During the period of dimer formation the electron diffraction patterns progressively become dominated by diffuse scattering (see diffraction patterns samples annealed between 300-400°C in Figure S6 in the Supporting Information). After 6 h at a temperature $T$ = 200°C, the fraction of TriAc molecules having formed dimers is 27±5%. As expected, annealing at a higher temperature $T$ = 300°C more efficiently promotes the formation of dimers: a 10 min annealing already results in 36±6% of the TriAc molecules forming dimers (Figure 1c). By estimating the percentage of TriAc dimers formed at these two temperatures, and assuming a $v_0 \exp(-E/k_BT)$, Arrhenius-like decay ($k_B$ being Boltzmann's constant) with a $v_0 = 10^{13}$ Hz attempt frequency, an activation barrier $E$ between 1.5 and 1.6 eV is derived.

The analysis of the STM data gives a 152±16° angle between the two connected legs of the dimer and a 1.85±0.30 nm distance between the central rings of the TriAc molecules (Figure 1d). The latter value fits well with the distance of 1.92 nm obtained from DFT calculations (Figure S7) for an unsupported stable structure consisting of two TriAc molecules having formed one covalent C=C double bond between the methyl group of the first TriAc molecule and the carbonyl group of the second TriAc molecule, releasing a water molecule in the process. This product corresponds to the first aldol condensation, i. e. the initiation of the polymerization process (Step 1 of Scheme 1).

By extending the annealing time, more branched structures are observed (Figure S8 in the Supporting Information). Some of these structures correspond to the products of Step 2 of Scheme 1, i.e. to the convergent triple aldol condensation reaction of three TriAcs yielding a new benzene ring. The formation of these structures, which are distant one from the other by a few nanometers in the explored experimental conditions, marks the beginning of the propagation of the 2D polymerization reaction. Increasing the annealing and deposition temperatures above 350°C more efficiently promotes this propagation than increasing the annealing time. Disordered nanoporous networks, whose coverage can reach 100% of the Au(111) surface depending on the amount of TriAcs evaporated, are achieved through the polymerization process (Figure 1e and Figure S9 in the Supporting Information). The pores of these networks are of various kinds (Figure 1f), in terms of shape, reticulation nodes, and progress of the reaction. The latter can be characterized by the fraction of TriAcs, one acetyl group of which at least has not participated in the aldol condensation reaction. This fraction decreases with increasing surface coverage (Figure 2), from 35-40% for 30% surface coverage (corresponding to network patches of extension of the order of 10 nm) to 15-20% for full-coverage, for similar annealing/deposition temperatures (~400°C). This decrease can be understood as a manifestation of the decrease of the amount of the network's edges upon increasing coverage. The fraction also decreases with annealing and deposition temperatures (Figure 2), for instance by 10%, down to 15-20%, when increasing the deposition temperature from room temperature to 200°C, while with similar high annealing temperatures. Such a decrease is consistent with the expected temperature-promoted surface diffusion of the TriAcs.

The nanoporous network is prominently composed of four- and six-sided pores, usually containing at least one new benzene ring as a reticulation node (Figure 1f-h), each corresponding to the completion of one triple aldol condensation. In some instances two and three benzene rings per pore are formed, for four- and six-sided pores respectively (Figure 1g,h), which corresponds to fully reacted pores, shown as the last step of Scheme 1. In other instances we observe reticulation nodes which are larger than benzene rings (shown as elongated hexagons in Figure 1f, for instance visible at the bottom left of the image in the case of four-sided pores). Such pores could actually be rings comprising eight carbon atoms, which would correspond to a dual aldol condensation between the ketone functions of two species of kind Step 1 shown in Scheme 1. The coexistence of pores of different shapes obviously causes strong distortions in the network, as seen by careful inspection of Figure 1f. The proportion of four- and six-sided pores varies with the coverage of the network. At low coverages (30%, network patches of extension of the order of 10 nm, e.g. point corresponding to 350°C deposition temperature and 450°C temperature in Figure 2), roughly 20% more six-sided pores are obtained, while for full-coverage cases 20% less six-sided pores are obtained. We ascribe this difference to a reduced mobility of the TriAcs prior to reaction at high coverages, which hinders the formation of the more energetically favorable (in view of steric and bond-distortion considerations) six-sided pores. We do not observe a substantial change in the proportion of four- and six-sided pores with annealing and deposition temperatures, presumably because when the polymerization reaction has significantly progressed (boundary between yellow and green domains in Figure 2), the surface mobility of the reactive species has already strongly decayed.

Other types of pores are observed as well, yet in minority, which correspond to situations when the 2D polymerization process is ended at Step 1 of Scheme 1 (see Figure S10 in the Supporting Information), presumably due to a limited mobility of the TriAc molecules and oligomers on the surface.[35] Competing reactions, involving the metallic surface (no competing reaction exist in solution), could also hinder the aldol condensation reaction to fully proceed, as was found in other systems.[36] Such reaction could promote partial graphitization through the formation of extra C-C bonds. Overall, the network is fully covalently bonded. We find that the number of TriAc molecules needed to achieve the 2D covalent carbon network is 40 % smaller than in the dense supramolecular network addressed before, implicating a loss of molecules by desorption.

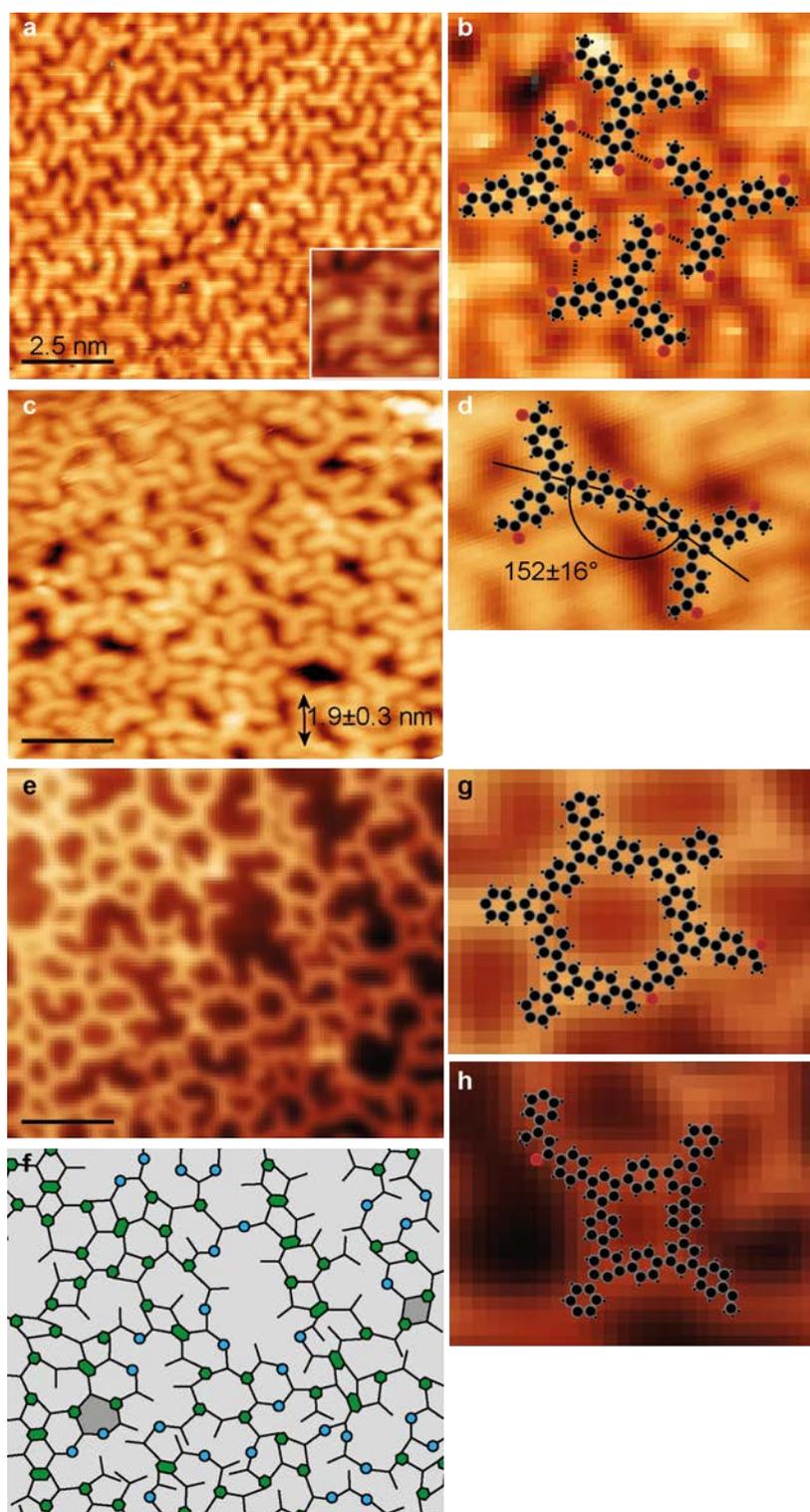

***Figure 1:*** *2D self-Assembly to a covalent carbon nanoporous network on Au(111), observed with STM. Inset: TriAc molecules exhibiting non uniform conductance after a tip change (a) TriAc molecules deposited at 200°C, leading to a H-bonded supramolecular network. (b) Zoom-in showing the arrangement of the supramolecular network with the hydrogen bonding highlighted by the dashed line. (c) At temperatures between 300 and 400°C, the system becomes disordered and the monomers begin to form TriAc dimers bonded together by covalent C=C bonds. (d) Zoom-in of a dimer. (e) 2D covalent nanoporous network. (f) Schematized view of the network in (e): the reacted TriAc molecules are shown as three-legged stars, green hexagon mark the position of extra carbon rings (see text for details), and disks mark the position of reticulation nods corresponding to*

*Step 1 in Scheme 1. (g,h) Zoom-in views with the presumed atomic arrangement superimposed onto the image, for six- and four-sided pores marked in dark grey in (f). Scale bares measure 2.5 nm. All STM topographs were obtained at +1.25 V and 0.1 nA.*

The 2D covalent carbon network is stable under UHV up to 500°C, a temperature from which the surface coverage decreases, due to desorption of some molecules, which implies the breaking of covalent bonds, presumably the C-C bonds attached to the remaining O atoms. From 550°C, the network undergoes strong modifications. We speculate that these modifications are related to those occurring during the thermal reaction between aromatic quinone molecules,[37] coronene,[38] or waste molecules,[39] and yielding graphene. Figure 2 summarizes our observations as a function of annealing and deposition temperature.

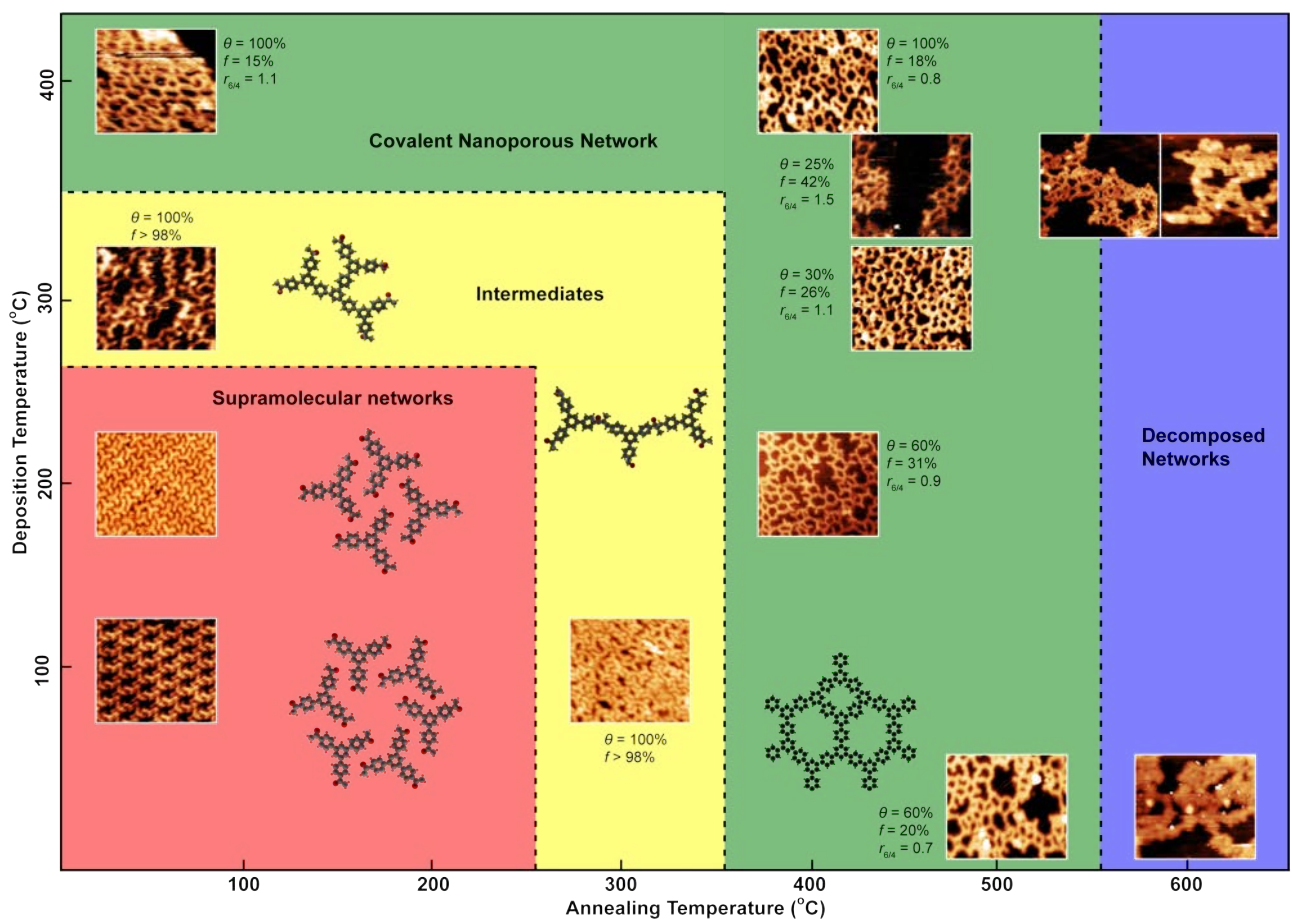

***Figure 2:*** *Diagram representing the four kinds of phases (red, yellow, green, and blue sectors of the diagram) encountered as a function of deposition and annealing temperatures. Instances of both low and high coverage ($\theta$) are shown. The fraction of TriAc molecules having at least one unreacted acetyl group (f) and the ratio in the number of six- to four-sided pours ($r_{6/4}$) are specified whenever relevant. Annealing time is 10 min. STM topographs are 20 nm-wide and were acquired at +1.25 V and 0.1 nA.*

DFT calculations were performed corresponding to a perfect (thus, ideal) lattice of six-sided pores after full completion of the triple aldol condensation, without the presence of the substrate (Figure 3a) and for the smallest coincidence lattice between an ordered 2D covalent carbon network on Au(111) (Figure 3b). These calculations predict that the benzene rings shall exhibit little distortion and lie flat on Au(111), while unsupported structures are distorted with a rotation along the axis of the intermediate benzene rings (Figure S11 in the Supporting Information). The average molecule substrate distance is predicted to be 0.31 nm, varying only by 4.5 % across the plane of the network. Such a distance is typical of a weakly bond system. As a comparison, graphene grown on gold, a system known for its weak interaction with the substrate, exhibits an experimentally determined separation distance of 0.33 nm.[40,41] The partial charge map (Figure S3 in the Supporting Information) integrated between Fermi level and 0.5 eV above is in good agreement with the STM observations, although the former is better spatially resolved, as is expected since the calculation does not include the effect of the STM tip.

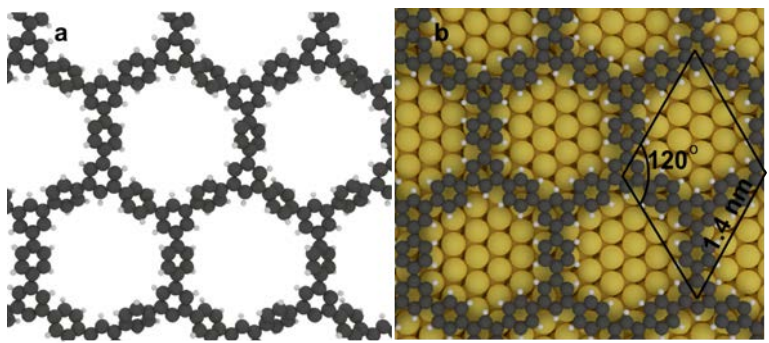

*Figure 3:* Top views of the structure of the periodic 2D covalent nanoporous network, inferred from DFT, without (a) and in the presence (b) of the Au(111) substrate.

No substantial changes are visible between the STM topographs obtained before and after exposure to air, even without annealing the sample after introducing it back into UHV (compare Figures 4a and b). The only feature that arises from exposing the sample to air is that the imaging conditions are less stable likely due to molecules from the atmosphere adsorbed on the sample, while the carbon framework remains intact. These observations establish the high stability of the 2D covalent carbon network, presumably a carbon backbone saturated with C-H bonds, to which molecules from air bond only weakly. This claim is further supported by Raman spectroscopy performed in air. Figure 5 shows the Raman spectra for a 2D network synthesized on Au(111) and for the TriAc monomers drop-casted onto Au(111) then allowed to dry. The spectra for the 2D network displays a peak centered at 1585 cm$^{-1}$, and a full-width at half maximum (FWHM) of 30 cm$^{-1}$. This peak is akin to the characteristic signature of a $sp^2$ hybridized carbon-conjugated systems, whose width would stem from disorder. In fact, this peak presumably corresponds to the so-called G vibration mode, which develops in largely extended $sp^2$ carbon systems. It differs from the mode observed in the case of the drop-casted monomers, which is centered at 1605 cm$^{-1}$ and as a 7 cm$^{-1}$ FWHM. Such features most probably relate to the stretching vibration mode of the carbon-carbon bonds in the monomer. Neither for the drop-casted monomer and the 2D network do we observe the characteristic features of the C=O stretching modes, which are expected to show up as peaks at about 1680 and 1820 cm$^{-1}$. Though they are indeed not expected for the reacted network, they are expected for the drop-casted monomers. In the latter case, their absence is presumably due to the strong and extended background in the Raman spectrum yielded by the gold substrate. On a quartz substrate on the contrary, we do observe a peak at ~1680 cm$^{-1}$ for the drop-casted monomers; however this substrate lacks the catalytic activity for synthesizing the reacted network. A perfect isolated 2D network is expected to give rise to a G mode with a narrower peak, appearing at higher wavenumber, than the one we observe. We interpret these observations as an effect of strain, which is known to modify the G peak position and width in a related system, graphene.[42] In this scenario, the G peak of the 2D network actually encloses contributions from regions with varying local strains, which coexist due to the presence of different pore geometries, to the coalescence of growing domains, and to the mismatch in thermal expansion coefficient between the network and Au(111). As such, the G mode width provides an indirect evidence of covalent bond formation. In this light the observed G mode for the reacted network shows strong similarities with that found in polymers.[43] The inset of Figure 5 shows a Raman map, displaying a signal for the peak centered at 1585 cm$^{-1}$ extending over 80% of the surface, indicating a large extension of the reacted network across the gold. These traits of sp$^2$ carbon found across nearly the entire surface, clearly indicate a fully-extended covalent network comprised of carbon. Overall, both molecular resolution imaging and vibrational spectroscopy indicate that the covalent bonds of the 2D networks are mostly unaffected by exposure to air, as is also the case in other 2D covalent networks.[44,45] Such stability is desirable in the prospect of applications outside UHV.

The triple aldol condensation reaction, performed with TriAc molecules, is a 2D polymerization process which is not directional in essence. Using templated surfaces (e.g. vicinal ones) or tailored monomers (e.g. with specific shape and/or reactivity), like was done for other kinds of surface reactions[46,44] could allow to achieve one-dimensional polymerized covalent networks, owing to lateral hindering of the monomers' mobility and reaction self-steering respectively.

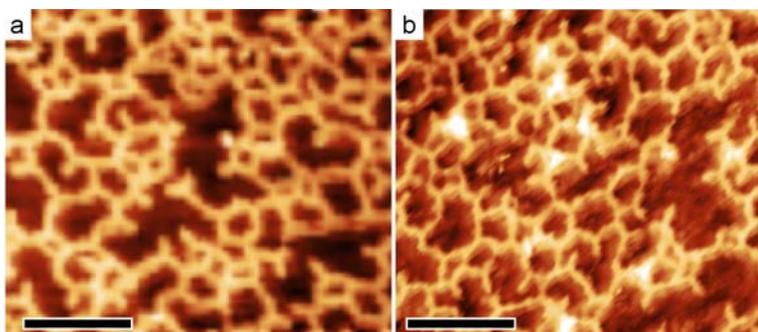

*Figure 4:* STM topograph of (a) a 2D covalent nanoporous network synthesized on Au(111) under UHV (deposition at 200°C; annealing at 400°C); and (b) of a network prepared under similar conditions, exposed to air, and introduced back in UHV. Images were acquired at +1.25 V and 0.1 nA.

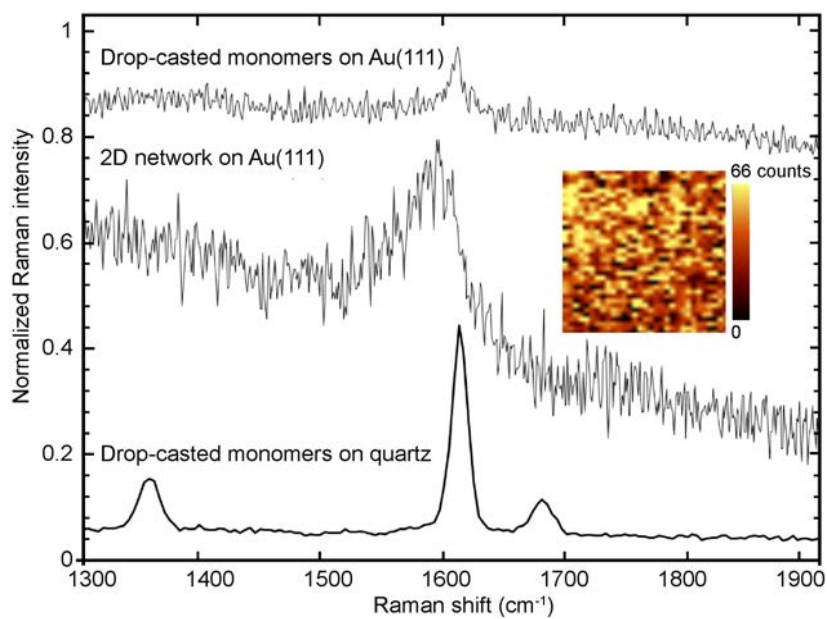

*Figure 5:* *Raman spectrum acquired in air for the 2D covalent nanoporous network on Au(111) (bottom curve) and the monomer drop-casted onto Au(111) (two top curves). The peak at 1590 cm$^{-1}$ is characteristic of a sp$^2$ carbon bonded systems. Spectra averaged over two 100 s acquisitions at the same position with a laser spot size of 1 μm. Inset: 30×30 μm$^2$ map of the peak intensity at 1590 cm$^{-1}$ for the 2D covalent nanoporous network.*

**Methods**

***Surface reaction of Triac molecules.*** TriAc molecules (melting point 252°C, 99.99% purity) were synthesized according to a procedure described in the Supporting Information. Surface reactions were performed in a UHV system with 10$^{-10}$ mbar base pressure. The molecules were evaporated with the same flux using two evaporator types, a multi-pocket commercial Kentax one and a home-made one, in which the quartz crucible containing the molecules is resistively heated, in our case at temperature of 200°C. The deposition rate of the TriAc molecules was calibrated with a quartz microbalance prior to deposition assuming a 1 g/cm$^3$ molecular density, and the deposition rate was set at 0.5 monolayers/min for a full coverage as obtained in Figure 1a. An aligned and polished Au(111) crystal with dimensions of 12×3 mm$^2$ and 0.5 mm thickness was purchased from Surface Preparation Laboratory. Prior to molecule deposition the Au(111) surface was prepared by repeating cycles of Ar$^+$ sputtering at 0.8 keV and annealing at 900 K until no evidence of carbon was detected, as verified by in situ Auger electron spectroscopy, and the herringbone reconstruction was fully developed, as verified by in situ STM.

***Characterizations.*** All STM measurements were performed under UHV and at room temperature, in the same system where the samples were prepared, using a commercial Omicron variable temperature STM/AFM, with electrochemically etched W tips. All images were obtained in constant current mode, with the tunnel bias applied to the tip. Images obtained for angles and distances were corrected from drift due to the piezoelectric elements, which was determined by tentatively recording two successive images at the same location. Micro-Raman spectroscopy was performed with a commercial Witec Alpha 500 spectrometer setup with a dual axis XY piezo stage in a back-scattering/reflection configuration. Grating used has 1800 lines/mm which confer a spectral resolution of 0.01 cm$^{-1}$ for 100 s integration time. Laser excitation wavelengths of 532 nm (solid state argon diode) was used. Raman spectra were recorded in air with a Nikon ×100 objective (NA = 0:9) focusing the light on a 320 nm diameter spot (532 nm light), and with a Mitutoyo ×50 objective (NA = 0:75). A background, whose intensity decays with increasing wave-number, corresponding to the luminescence from the Au substrate, is observed in Figure 6.

***Density functional theory calculations.*** DFT calculations were performed using the VASP code, with the projector augmented wave (PAW) approach.[47] The exchange correlation interaction is treated within the general gradient approximation parameterized by Perdew, Burke and Ernzerhof (PBE).[48] When the gold substrate was taken into account, long range dispersion corrections (van der Waals interactions) were accounted for by using Grimme corrections.[49] The dispersion coefficient C6 and van der Waals radius R0 for Au are not listed in the original paper by Grimme and thus have been taken from Amft *et al.*[50] These values for C6 and R0 are 40.62 J×nm/mol and 1.772 Å respectively. The Au(111) surface was modelled using a four-layer slab that was first relaxed. The TriAc covalent network was then added. Gold atoms were kept fixed in the bottom layers, all other atoms were allowed to relax. The lateral periodicity was imposed by the smallest – commensurate – supercell chosen to accommodate a TriAc molecule, therefore a 3√3×3√3 periodicity of the Au(111) surface. Note the DFT calculations were performed for unreconstructed Au(111). The empty space in the direction perpendicular to the surface was chosen equal to 14 Å. After convergence, residual forces were lower than 0.02 eV/Å. Calculations of the isolated TriAc or TriAc dimer were performed in very large supercells to avoid spurious interaction due to periodicity.

**Supporting Information**
Synthesis details of the TriAc monomer, additional surface characterization and DFT calculations are provided in the Supporting Information.

**Acknowledgement**
JL acknowledges financial support for a research fellowship from the Nanosciences Foundations. We acknowledge Dr. S. Vlaic, Dr. Z. Han and Prof. Dr. F. Palmino for valuable discussions. JL is grateful to Prof. Kocabas for providing the Raman spectra for graphene grown on gold, displayed in the supporting information.[51]